\documentclass[twocolumn,prl,showpacs]{revtex4}
\usepackage{graphics}
\usepackage{graphicx}
\usepackage{pstricks}
\usepackage{amsmath}
\usepackage{amssymb}
\usepackage{pst-node}
\usepackage{multido}
\usepackage{pst-plot}
\usepackage{pst-coil}
\usepackage{psfrag}
\usepackage{verbatim}
\usepackage{pifont}
\usepackage{lscape}
\usepackage{bm}
\usepackage{enumerate}

\newcommand{\bra}[1]{\ensuremath{\langle#1|}}
\newcommand{\ket}[1]{\ensuremath{|#1\rangle}}

\newcommand{\be}{\begin{equation}}
\newcommand{\ee}{\end{equation}}
\newcommand{\bea}{\begin{eqnarray}}
\newcommand{\eea}{\end{eqnarray}}
\newcommand{\tr}{\mathrm{tr}}

\newcommand{\diag}{\textrm{diag}}

\newcommand{\dg}{\dagger}
\renewcommand{\S}{\mathbf{S}}

\newcommand{\vs}{\vspace{.1cm}}
\newcommand{\T}{\mathrm{T}}
\newcommand{\unisa}{Dipartimento di Ingegneria Industriale,
Universit\`a degli Studi di Salerno, Via Ponte don Melillo,
I-84084 Fisciano (SA), Italy; CNR-SPIN, Unit\`a di Salerno;
CNISM, Unit\`a di Salerno; and INFN Sezione di Napoli,
Gruppo collegato di Salerno, I-84084 Fisciano (SA), Italy}

\begin{document}

\title{Characterizing and quantifying frustration in quantum many-body systems}
\author{S.~M.~Giampaolo}
\affiliation{\unisa}
%%\affiliation{\order}
\author{G.~Gualdi}
\affiliation{\unisa}
%%\affiliation{\order}
\author{A.~Monras}
\affiliation{\unisa}
%%\affiliation{\order}
\author{F.~Illuminati}
\thanks{Corresponding author: illuminati@sa.infn.it}
\affiliation{\unisa}
%%\affiliation{\order}

\date{December 23, 2011}

\begin{abstract}
We present a general scheme for the study of frustration in quantum systems. We introduce a universal measure of frustration for arbitrary quantum systems and we relate it to a class of entanglement monotones via an exact inequality. If all the (pure) ground states of a given Hamiltonian saturate the inequality, then the system is said to be inequality saturating.
We introduce sufficient conditions for a quantum spin system to be inequality saturating and confirm them with extensive numerical tests. These conditions provide a generalization to the quantum domain of the Toulouse criteria for classical frustration-free systems. The models satisfying these conditions can be reasonably identified as  geometrically unfrustrated and  subject to frustration of purely quantum origin. Our results therefore establish a unified framework for studying the intertwining of geometric and quantum contributions to frustration.
\end{abstract}

\pacs{03.67.Mn, 03.65.Ud, 05.50.+q, 75.10.Jm}

\maketitle

Frustration consists in the impossibility
of determining configurations that minimize simultaneously the energy of competing
interactions \cite{Toulouse1977,Villain1977,Binder1986}. In recent years it has been realized that classically unfrustrated systems can have frustrated quantum counterparts \cite{Wolf2003,Nielsen2004,Giampaolo2010,Eisert2010}. Indeed, in the quantum case, additional sources of frustration may arise due to non-commutativity and entanglement \cite{ Nielsen2004,Acin2007}. Hence,
although the notion of frustration has often been considered from the perspective of statistical physics, in the quantum domain interesting novel phenomena take place even in the presence of few entangled elements \cite{Acin2007,Nielsen2004,Facchi2010}. It is however far from clear how to distinguish geometric from purely quantum frustration and whether the distinction is free of ambiguities. To the best of our knowledge this issue has never been addressed. Moreover, despite frustration being a well-defined and intuitive concept, a measure for quantum systems which fully captures all its aspects is still lacking.
Existing proposals for quantifying frustration in quantum systems focus on particular aspects of the phenomenon. Some proposals are based on classical equivalents of a given quantum system \cite{Lacorre}, making it impossible to recover information about quantum correlations. Others apply only
in the thermodynamic limit \cite{Balents,Ramirez}, or on the competition between local and nonlocal interactions from a purely quantum perspective \cite{Nielsen2004}, thus making it challenging either to study finite-size systems or to understand the transition to classically frustrated systems.
The need for a systematic investigation of this issue at a foundational level is thus quite compelling. Even more so, as quantum simulators of classical as well as quantum frustrated magnetic systems, at least in the simplest Ising and $J_1$-$J_2$ cases, are being demonstrated with ultracold atoms in optical lattices \cite{Sengstock,Greiner} and trapped ions \cite{Monroe}, or proposed, e.g. with cold Coulomb crystals \cite{Plenio}. Furthermore, such an investigation would be of great help and guidance for the design of entanglement-based numerical simulations of frustrated quantum spin models \cite{Verstraete}.

The aim of this work is to introduce a universal measure of frustration for quantum systems and define a unified framework suitable to understand the intertwining of the geometric and quantum contributions to frustration. To this end we focus on the microscopic properties of finite-size models from a purely quantum perspective without resorting to semiclassical approximations. After introducing a universal measure of frustration, we prove that it is an upper bound to a class of entanglement monotones that in particular cases reduce to the bipartite geometric entanglement. We then establish sufficient conditions for a quantum spin system to saturate the bound and support them with extensive strong numerical evidence. Finally, we show how these conditions essentially generalize to the quantum domain the Toulouse criterion for frustration-free classical systems.

Consider a system in a pure state $\rho=\ket{\psi}\bra{\psi}$, and let $\Pi$ be the projector onto a given subspace. Then
\begin{equation}\label{f}
	f=1-\tr[\rho \Pi],
\end{equation}
quantifies how much $\rho$ fails to fully overlap with the subspace selected by $\Pi$. Let now $\rho=\ket{G}\bra{G}$ be the ground state (GS) of a many-body system $H = \sum_{S}h_{S}$ and $\Pi_S$ the projector onto the GS of the local interactions $h_{S}$ corresponding to subsystem $S$. Then $f_{S}=1-\tr[\rho\Pi_S]$ is a well-defined and unambiguous measure of the frustration of $h_{S}$. On the other hand, denoting  by $R$ the rest of the system, consider the following entanglement monotone:
\begin{equation}\label{mon}	
	E^{(d)}(S|R)=1-\sum_{i=1}^d\lambda_i^\downarrow(\rho_{S}) \; ,
\end{equation}
where $\lambda_i^\downarrow(\rho_{S})$ are the eigenvalues of $\rho_{S}=\tr_{R}\rho$ in decreasing order and $d<\mbox{dim}[\mathcal H_S]$, with $\mathcal H_S$ the Hilbert space of $S$. Notice that $E^{(d)}(S|R)=E^{(d)}(R|S)$. The r.h.s. of Eq. \eqref{mon} vanishes only on states with Schmidt rank smaller or equal than $d$~\cite{Turgut}; for $d=1$ it reduces to the bipartite geometric entanglement, defined as the distance from the set of bi-separable states \cite{Dellanno}.
Fix now $d$ to be the degeneracy of the local interaction $h_{S}$ (i.e. $\Pi$ is rank-$d$). Then, by Cauchy interlacing theorem (See also Lemma 1 in the supplementary material) it follows that
\begin{equation}\label{eq:inequality}
	f_{S}\geq E^{(d)}_{S} \; ,
\end{equation}
where $E^{(d)}_{S} \equiv E^{(d)}(S|R)$ is the distance from $\ket{G}$ to the closest state with Schmidt rank $r\leq d$~\cite{Turgut}. %On the other hand, $f_{S}$ is the distance between $\ket{G}$ and the closest state minimizing $h_{S}$, $f_{S}=\min_{\Pi\ket\Phi=\ket\Phi}(1-|\langle G|\Phi\rangle|^2)$. Realizing that any state minimizing $h_{S}$ has Schmidt rank at most $d$ provides an intuitive geometric representation of Eq.~\eqref{eq:inequality}.

Despite its apparent simplicity,  Eq.~\eqref{eq:inequality} has the remarkable feature of directly relating frustration to entanglement. This quantitative relation holds for {\it any} pure state $\rho$ and {\it any} interacting quantum system and hence is universal. Actually, Eq.~\eqref{eq:inequality} holds as well for any mixed state, although in this case $E_S^{(d)}$ is no longer an entanglement monotone. An immediate consequence of Eq. \eqref{eq:inequality} is that the frustration-free condition, $f_{S}=0 \forall S$ is a bound on the maximum Schmidt rank of the global ground state $\ket{G}$. For interactions $h_S$ with nondegenerate local GSs $\ket{G_{S}}$, $d=1 \forall S$, this implies the separability of the global GS in the tensor product of the local GSs: $\ket{G} = \bigotimes_{S}\ket{G_{S}}$. That the absence of frustration should be related to some form of factorization of the GS had already been observed, at a semi-quantitative level, for models in transverse fields \cite{Giampaolo2010}.
On the other hand, saturation of the inequality, $f_{S}=E^{(d)}_{S}$ for some $S$, imposes a block-diagonal form of the reduced state $\rho_{S}$, with eigenvalues $\lambda_i^\downarrow(\rho), i=1,\ldots,d$ corresponding to the block spanned by $\Pi$ (see Lemma 1 in the supplementary material). Hence $\rho_{S}$ cannot exhibit coherence between the lowest and excited energy levels of $h_{S}$: the largest contribution to $\rho_{S}$ must come from the local ground subspace.

Summarizing the above discussion, we can state that a GS $\rho = \ket{G}\bra{G}$ of a many-body Hamiltonian
of the form $H=\sum_{S}h_{S}$ is a {\em frustration-free} (FF) state if and only if  $f_{S}=0\; \forall S$, and is an {\em inequality saturating} (INES) state if and only if  $f_{S}=E^{(d)}_{S}\; \forall S$. Clearly, a FF state is also an INES state. A Hamiltonian $H$ is then a \emph{FF Hamiltonian} if all its GSs are FF states, and is an {\em INES Hamiltonian} if all its GSs are INES states. In general, it is easy to show that states with at least one $f_{S}>0$ have higher energy compared to the corresponding FF state. Hence, if a model is globally degenerate and one global GS is a FF state, then this is true for all other GSs.
Unlike the FF property, the INES property is not universal, {\em i.e.} it does not necessarily apply to all the GSs of a system. This is due to the fact that unlike the FF property, the INES property does not specify the GS energy. Total frustration $F$ can then be defined out of the set of local measures $f_{S}$'s by arithmetic or weighted averaging.

We will now exploit Eq.~\eqref{eq:inequality} to generalize the classical criteria for the absence or presence of frustration to the quantum domain and to understand the intertwining of geometric and purely quantum contributions to frustration. According to Toulouse [Formulation 1] \emph{A classical Hamiltonian $H$ is frustrated if and only if it is impossible to transform $H$ into a fully ferromagnetic Hamiltonian by means of local spin inversions}. This occurs only when a closed loop exists with an odd number of antiferromagnetic interactions~\cite{Toulouse1977,Villain1977,Wolf2003}. Indeed, the Toulouse criterion computes exactly the parity of antiferromagnetic bonds on a closed loop, according to: [Formulation 2] \emph{For a given Hamiltonian a loop is frustrated if the quantity $P=(-1)^{N_{af}}=-1$, where $N_{af}$ is the number of antiferromagnetic bonds}. %Both  formulations  are deeply related to the classical nature of the systems which makes the generalization to the quantum case not straightforward. %{\bf Ising Model}
Next, consider the simplest quantum extension of classical models, the quantum Ising model $H=\sum_{i<j}J_{ij}S^z_iS^z_j$ ($d=2$). Clearly, the model is unfrustrated if the GS is of the form
$\ket\psi=\bigotimes_{i=1}^N(S^x_i)^{\gamma_i}(\alpha \ket{\!\!\uparrow\uparrow\!\cdots}+\beta\ket{\!\!\downarrow\downarrow\!\cdots})$, with $\gamma_i=0,1$.
Indeed, in this case $f_{ij}=E^{(2)}_{ij}=0$ $\forall S \equiv ij$, and Eq.~\eqref{eq:inequality} is saturated by all GSs.
On the contrary, if the model is frustrated there exists at least one GS which is not an INES state. In fact,  all separable GSs exhibit $E_{ij}^{(2)} = 0$ on all pairs and $f_{ij} > 0$ on at least one pair.
%Indeed, in this case, all GSs exhibit $f_{ij}>0$ for some $ij$. Thus all separable GSs, being $E^{(2)}_{ij}=0$ on all pairs,
%will not be INES states.
Therefore, in terms of Eq.~\eqref{eq:inequality}, the immediate extension to the quantum Ising case of the Toulouse condition is [Formulation~3]: {\em A quantum Ising Hamiltonian is frustrated if and only if it is not an INES model}. Although the Ising model does not contain quantum features, such as non-commutativity of the local interactions, and therefore can only exhibit geometric frustration, the possibility of restating Toulouse criterion in terms of $f$ and an entanglement monotone $E_{S}^{(2)}$ is remarkable as it provides the first bridge between the classical and the quantum domains.

From the Ising example we learn that in the quantum domain the relevant information to detect frustration is the existence of GSs not being INES states, rather than that of GSs not being FF states. This is because for quantum systems frustration arises non only from topological constraints (so-called \emph{geometric frustration}) but as well from purely quantum ones.
Consider the following classically unfrustrated Heisenberg Hamiltonian for four spins on an open chain: $H=\sum_{i}{ \S_{i}}\cdot{ \S_{i+1}}$, where ${\S_i}=(S^x_i,S^y_i, S^z_i)$, $d=1$ and the global GS of $H$ is non degenerate. The quantum version of this model is frustrated, as in the GS the total measure of frustration $F=N^{-1}\sum_{ij}f_{ij}=\frac{1}{6}(3-\sqrt3)$, where $N=3$ is the number of bonds. Remarkably, the frustrated GS is an INES state, hence the frustrated model is still an INES model, in contrast to what occurs in the Ising case. Let us now add to $H$ some geometric frustration {\em i.e.} $H'=H+\sum_{i}{ \S_{i}}\cdot{ \S_{i+2}}$. Now one has $f_{ij} = 2/5 > E^{(1)}_{ij} = 1/3 \forall (ij)$, and hence the frustrated GS is not an INES state. Indeed, the Ising model can only exhibit geometric frustration, even in its quantum version, as the local GSs are always unentangled.
%%In fact, due to the degeneracy of the local interactions, no quantum correlations are imposed on the local %%GSs.  %This model can thus not be frustrated {\em and}  an INES model as, if it is an INES model, then it must %%also be a FF model.
On the contrary, the GS of each Heisenberg local pair term is a maximally entangled Bell state. Accordingly, the model can either be an INES model or not, depending on the topology of the system. Actually, even if geometric frustration is absent, yet monogamy of entanglement \cite{CKW} prevents the global GS from minimizing all the local terms, and the model is both frustrated and INES. This strongly suggests that failure to saturate Eq.~\eqref{eq:inequality} is a signature of the presence of geometric frustration.

% Although  the set of $f_{ij}$'s provides us with a conceptually simple measure of frustration, nevertheless its computation requires the highly non trivial knowledge of the full GS of a many body system.
From the above discussion it follows that it would be highly desirable to identify conditions detecting {\em a priori} the nature of the frustration in a given quantum system.
Considering for the moment being only models with non-degenerate global GS, we approach the problem by observing that the classical Toulouse condition contains two main ingredients: a ferromagnetic model, which serves as the {\emph {prototype}} of  FF models, and a gauge group under which the FF property is preserved. We thus look for a prototype INES model and a gauge group under which the INES property is preserved. To fix the stage, consider the
general $XYZ$ (Heisenberg) exchange Hamiltonian
\begin{equation}\label{heisham}
H\!=\!\sum_{ij}h_{ij}\!=-\sum_{ij,\mu}\! J^\mu_{ij} S_i^\mu S_j^\mu,
\end{equation}
where $\mu=x,y,z$, with coupling vectors $\vec J_{ij}\!=\!(J^x_{ij},J^y_{ij},J^z_{ij})$ on arbitrary graph geometries. \\
\noindent[{\bf Prototype model}]: A non-degenerate quantum Hamiltonian as in Eq.~\eqref{heisham} will be called {\em prototype} if 1a) there exists at least one local GS common to all local pair interactions $h_{ij}$'s, and 1b) each local coupling vector $\vec J_{ij}$ has non-negative components (i.e. it is a full ferromagnet: $J_{ij}^\mu\geq 0~\forall i,j,\mu$ and there exists a common two-body state vector $\ket\phi$ which is ground state of all $h_{ij}$).

\noindent[{\bf Conjecture 1}]: {\em All prototype models are INES.}

Note the remarkable parallelism between condition {\em 1a)} and the FF condition, which essentially requires the existence of at least one {\em global} GS common to all $h_{ij}$'s.
From our conjecture, it follows that if $d=1$ then all $h_{ij}$'s must admit the same Bell state as GS. Since the GS of a given $h_{ij}$ with positive $\vec J_{ij}$ is determined by the lowest value component of the vector, it follows that it must be along the same axis for all $\vec J_{ij}$ in the prototype model. Any model obtained from the prototype by means of tensoring local unitary operations is clearly still an INES model with the same set of $f_{ij}$ and $E^{(d)}_{ij}$'s. In fact, one has

\noindent [{\bf Conjecture 2}]: {\em Any model $H'=H^{\T_K}$ which can be obtained from a prototype model $H$ by partial transposition on any set of sites $K$ is still an INES model}.

Applying partial transposition~\cite{Partialtranspose} to a prototype Hamiltonian changes $f_{ij}$ and $E^{(d)}_{ij}$; however, saturation of Eq.~\eqref{eq:inequality} is preserved. Although an analytical proof of these conjectures appears challenging, in the supplementary material we provide compelling numerical evidence of their validity. This is a very significant result, as it implies that not only local rotations but also partial transpositions do preserve the INES property. It is not {\it a priori} obvious that this would be the case, since, as already noted, both sides of inequality \eqref{eq:inequality} are typically changed by a partial transposition. Indeed, partial transposition is intimately related to parity, as
%\begin{equation}
$	P_k(H)=S^y_k H^{\T_k}S^y_k $,
%\end{equation}
where $P_k$ is the parity transformation on site $k$ [$P( \S)=- \S$]. Thus, according to Conjecture 2, the local gauge group is $G=SU(2)\otimes\mathbb{Z}_2$. An element of
$SU(2)$ acting on $S^\mu$ is represented by a transformation $R\in SO(3)$, whereas a parity transformation is simply $-\openone$. Hence, spin operators transform according to the $O(3)$ representation of $G$, and a local gauge transformation $\mathfrak g=\bigotimes_i g_i$ (with $g_i\in G$) maps two-body interactions $\sum_{\mu\nu} J_{ij}^{\mu\nu}S_i^\mu S_j^\nu$ into $\sum_{\mu\nu}[R_i^\top J_{ij}R_j]^{\mu\nu}S_i^\mu S_j^\nu$.
%%We stress that the subscripts $i$ and $j$ are only site bookkeeping labels and should not be regarded as %%matrix or vector indices. The latter are indicated by Greek indices and will be omitted when no confusion %%arises.
Given a general $XYZ$ Hamiltonian $H=-\sum_{ij}\sum_\mu J_{ij}^\mu S^\mu_iS_j^\mu$, we derive necessary and sufficient conditions for $H$ to be equivalent to a prototype model under the action of some $\mathfrak g\in G^{\otimes N}$. Consider two sites $a$ and $b$ and let \mbox{$p(a\rightarrow b)=$}$\,\{(a,i_1),(i_2,i_3),\ldots,(i_k,b)\}$ be any path from $a$ to $b$, where all pairs $(i,j)\in p$ interact. Define the {\em sign of the path $p$} as
\begin{equation}
\pi(p(a\rightarrow b))=\prod_{(i,j)\in p}  \diag({\rm s}^x_{ij},{\rm s}^y_{ij},{\rm s}^z_{ij}) \; ,
%	\left(
%		{\small \begin{array}{ccc}
%			{\rm sign}(J^x_{ij}) & 0 & 0 \\
%			0 & {\rm sign}(J^y_{ij}) & 0 \\
%			0 & 0 & {\rm sign}(J^z_{ij})
%		\end{array}}
%	\right)
\end{equation}
where ${\rm s}^\mu_{ij}={\rm sign}(J^{\mu\mu}_{ij})$ and we define ${\rm sign}(0)=1$, with the product
taken over all adjacent pairs $ij$ belonging to the path $p(a\rightarrow b)$. Note that $\pi(p(a\rightarrow   b))=\pi(p(b\rightarrow   a))$. We state the following

\noindent[{\bf Theorem}] {\em Necessary and sufficient conditions for a Heisenberg Hamiltonian $H$ to be mapped into a prototype model by a local gauge transformation are
\begin{enumerate}[2a)]
	\item All coupling vectors $\vec J_{ij}$,  have the smallest absolute value component along the same axis;
	\item For any pair of spins $a$ and $b$, $\pi(p(a\rightarrow   b))$ is independent of the path $p$ from $a$ to $b$.
\end{enumerate}}
As shown in the supplementary material, condition {\em 2b)} guarantees that the system can be brought to a fully ferromagnetic system. When this condition is met, the dependence of $\pi$ on the path can be dropped. Then one can consider $\pi(a\rightarrow b)$ as a ``conservative field'' which dictates the local transformation that has to be applied in $b$ so that the sign of any path from $a$ to $b$ is positive. If this holds for all $b$, one can turn all the couplings to positive by means of local transformations. On the other hand, condition {\it 2a)} is related to the existence of the same Bell state as the common ground state of all the two-body interactions. A remarkable simplification occurs when the system is translationally invariant. In that case:

\noindent{[{\bf Theorem}] \em If a model satisfies
%\begin{equation}\label{ToulouseVec}
$\pi(\ell)=+\bm{\openone} \;$,
%\end{equation}
for every loop $\ell$ in the elementary cell and all coupling vectors have the smallest component along the same axis, then it is an INES model} (Proof in the supplementary material).
This Theorem encodes in compact form three Toulouse's criteria, one for each spatial direction. This strongly suggests that the class of INES models defined by conditions {\em 2a-b)} can be identified as that of geometric frustration-free spin-1/2 quantum models.
%%The fact that the conditions {\em 2a-b)} are sufficient but not necessary highlights the qualitative nature %%of Toulouse's criterion. As not all classical frustrated spin models present certain features (e.g. %%nonvanishing zero-point entropy), failure to saturate Eq.~\eqref{eq:inequality} is also restricted to %%certain quantum models with geometric frustration.
Our analysis reveals that the quantum nature of the model affects the very notion of geometric frustration. Condition {\em 1a)} or equivalently {\em 2a)} is a consequence of the existence of three inequivalent ferromagnetic states, {\em i.e.} the triplet states. This constraint, however, is only relevant in systems with inhomogeneous couplings. Otherwise, conditions {\em 2a-b)} simply reduce to a generalized form of Toulouse's criterion. Being the class of geometric frustration-free quantum systems more restrictive than its classical analogue, one may expect that further investigation reveals generic properties which so far failed to be properly generalized.

On the other hand, the class of geometric unfrustrated quantum systems is strictly larger than that of frustration-free systems. A deeper investigation of geometric FF quantum models might thus unveil several relevant applications. For example, being the largest eigenvalues of $\rho_{ij}$ identified by $\Pi_{ij}$ may provide guidelines for designing optimal entanglement renormalization algorithms and other tensor network ansatz~\cite{MERA}. Moreover, the measure of frustration that we have introduced and its relation to geometric entanglement is not restricted to local two-body Hamiltonians or spin-1/2 systems and might be exploited to gain understanding of geometric frustration in arbitrary quantum many-body systems.
%%The natural relation between the frustration measures $f_{ij}$ and the entanglement monotones %%$E^{(d)}_{ij}$ reveals a strong tie between the two physical quantities.
%%For two-local spin-1/2 Hamiltonians it is tempting to define the {\em entanglement deficit} %%$\Delta=f-E^{(d)}$ as a measure of geometric frustration. However, the existence of INES models that fail to %%satisfy conditions {\em 2a-b)} prevents us from doing so. These generally fail to remain INES models under %%local parity transformations. It would hence be interesting to see if these models behave differently from %%other geometrically frustrated ones.

According to our results,  the INES class of models is larger than that of geometric FF, thus suggesting that the INES property has deeper implications than merely detecting the presence or absence of geometrical frustration. For example, the fact that  INES geometrically frustrated models  may behave differently from non INES  geometrically frustrated ones hints at a significant role of the INES property as a diagnostic tool for quantum phase transitions in complex and frustrated models.
Indeed, consider a frustrated system such as the elementary cell of the pyrochlore lattice $H=J( \S_1\cdot \S_2+ \S_3\cdot  \S_4)+( \S_1+ \S_2)\cdot( \S_3+ \S_4)$~\cite{Pyrochlore}. For $J>1$ the system is INES since it is in a dimer phase and all dimers have $f=E^{(1)}=0$, whereas all other bonds are in a maximally mixed state, hence $f=E^{(1)}=3/4$. Exact diagonalization shows that the transition from the dimer ($J>1$) to the plaquette phase ($0<J<1$) corresponds to the transition from INES to non-INES, with $f_{12}=f_{34}=1>E_{12}^{(1)}=E_{34}^{(1)}=2/3$. This suggests a correspondence between the different quantum phases of a frustrated model and its INES or non-INES character. Indeed, the relation between geometric frustration and exotic matter phases has been previously pointed out in~\cite{Balents}. One may also ask how the presence/absence of geometric frustration and the INES/non-INES nature of a system gets reflected in computational~\cite{Vidal2010}, information-theoretic~\cite{Lacorre,Wolf2008} or thermodynamic terms~\cite{Ramirez}. Necessary steps toward further investigation of all these open fundamental questions will require achieving a rigorous mathematical control and understanding of our quantum Toulouse conditions as well as the generalization of our approach to globally degenerate systems. Although some instances of degeneracy can be easily accommodated within our scheme (e.g., odd number of spins), some others (e.g. thermodynamic degeneracy) are still elusive, but in very simple cases such as the $XY$ model.

The essential and intriguing role played by partial transposition cannot go unnoticed. The fact that the prototype models preserve their INES character under partial transposition is to be expected if INES has anything to do with the presence or absence of geometric frustration. Nevertheless, the preservation of the INES property is far from trivial because, unlike with the case of local unitary transformations, the GSs of a Hamiltonian {\em before} and {\em after} partial transposition are not immediately related. Indeed, we expect that the rigorous proof of our conjectures will shed further light on the role played by partial transposition and its relation to geometric frustration. From a directly physical point of view, as already noticed, our quantitative analysis confirms previous evidence \cite{Giampaolo2010} that some form of factorization of the global GS is a necessary ingredient in the characterization of FF systems. It is tempting to speculate that investigating the intertwining between frustration and factorizability of higher order ($k$-separability, including dimerization and trimerization) could lead to a unified framework for the
understanding of the relations between frustration, the role of hierarchical geometric entanglement \cite{Dellanno} in collective quantum phenomena, and the characterization
of entanglement and $k$-separability by local unitaries \cite{Monras2011}.

We acknowledge financial support from the EU STREP Project HIP, Grant Agreement n. 221889.

\appendix

\section{Supplementary Material}

{\bf Lemma 1:} {\em Let $\Pi$ be a $d$-dimensional projector, $\bar \Pi=\openone-\Pi$ its complement and $\rho$ a density operator. Then, saturation of the inequality $\tr[\rho \Pi]\leq \sum_{i=1}^d \lambda^\downarrow_i(\rho)$ implies $\rho={\rm  diag}(p_1,p_2,\ldots p_d) \oplus \,\bar\Pi\rho\bar \Pi$ where $p_i=\lambda^\downarrow_i(\rho)$ are the eigenvalues of $\rho$ in decreasing order.}\vs

{\em Proof:} Let $A$ be any Hermitian operator and $\lambda^\downarrow_k(A)$ be the eigenvalues of $A$ arranged in decreasing order. From Cauchy's interlacing theorem~\cite{Bhatia} (CIT) we have
\begin{equation}
	 \lambda^\downarrow_i(\Pi\rho\Pi)\leq\lambda^\downarrow_i(\rho).
\end{equation}
The condition $\tr[\rho \Pi]=\sum_{i=1}^d \lambda^\downarrow_i(\rho)$ implies
\begin{equation}
	\sum_{i=1}^d \lambda^\downarrow_i(\Pi\rho\Pi)=\sum_{i=1}^d \lambda^\downarrow_i(\rho)
\end{equation}
which can only be satisfied if $\lambda^\downarrow_i(\Pi\rho\Pi)=\lambda^\downarrow_i(\rho)$ for all $i=1,\ldots,d$. This proves
\begin{equation}
	\rho_\Pi=\Pi\rho\Pi=\left(\begin{array}{ccc}\lambda^\downarrow_1(\rho) & &  \\ & \lambda^\downarrow_2(\rho) &  \\ &  & \ddots\end{array}\right)
\end{equation}
and $\rho$ has the block structure
\begin{equation}
	\rho=\left(\begin{array}{cc}\rho_\Pi & A \\A^\dg & \bar\Pi\rho\bar\Pi\end{array}\right).
\end{equation}
Since $\sum_{j=1}^d\lambda_j^\downarrow(\rho)=\sum_{j=1}^d \rho_{jj}$, we have that $\rho$ must be block-diagonal \cite{Markham}; hence:
\begin{equation}
	\rho=\left(\begin{array}{cc}\rho_\Pi &  \\ & \bar\Pi\rho\bar\Pi\end{array}\right).
\end{equation}
\hfill$\Box$

\section{Proofs of theorems}\label{app:proofs}
\noindent[Theorem] {\em Necessary and sufficient conditions for a Heisenberg Hamiltonian $H$ to be mapped into a prototype model by a local gauge transformation are
\begin{enumerate}[2a)]
	\item All coupling vectors $\vec J_{ij}$, have the smallest absolute value component along the same axis;
	\item For any pair of spins $a$ and $b$, $\pi(p(a\rightarrow   b))$ is independent of the path $p$ from $a$ to $b$.
\end{enumerate}}

When the second condition is met, the dependence of $\pi$ on the path can be dropped. Moreover, we have \mbox{$\pi(a\rightarrow   c)=\pi(a\rightarrow   b)\pi(b\rightarrow   c)$}. The conditions are clearly satisfied for the prototype model. The subgroup $H$ of $G$ which preserves the Heisenberg structure of the Hamiltonian when acting on individual sites is generated by the operations $P^\mu(\cdot)=P(\S^\mu\cdot\S^\mu)$ which all have diagonal $O(3)$ representations. Notice that each $P^\mu$  combines of an $SU(2)$ transformation and a parity inversion. The representation of $P^\mu$ is given by $R^\mu=\diag((-1)^{\delta_{x\mu}},(-1)^{\delta_{y\mu}},(-1)^{\delta_{z\mu}})$. Under these transformations the component in $\vec J$ with smallest absolute value cannot be changed without $H$ losing the Heisenberg form. Then, the path-independence of $\pi(a\rightarrow b)$ is clear from the fact that if $P^\mu$ acts either on $a$ or on $b$ it will flip the sign of {\em all} paths from $a$ to $b$. On the other hand, if $P^\mu$ acts on any other site, it will change none. This shows that {\em 2a-b)} are necessary conditions. To prove that they are sufficient, start from any spin $a$ and apply to any other spin $b$ the operation
$R_b=\pi(a\rightarrow b)$. Then, for any pair of interacting spins $i,j$, the resulting new couplings are given by $J'_{ij}=R_i^\top J_{ij}R_j=|J_{ij}|$, where $|J_{ij}|=\diag(|J^x_{ij}|,|J^y_{ij}|,|J^z_{ij}|)$. This ensures that the axis of minimum coupling is preserved under such transformations, and that all interactions are ferromagnetic. This proves the theorem.\hfill$\Box$
\begin{figure}[t]
	\includegraphics[width=.2\textwidth]{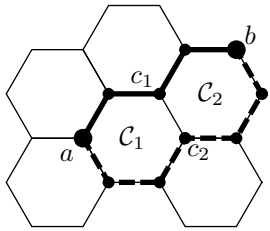}
	\caption{\label{fig:lattice}Two paths $p_1$ and $p_2$ going from $a$ to $b$, passing through $c_1$ [solid] and $c_2$ [dashed] respectively.}
\end{figure}

Testing conditions {\em (2a-b)} is particularly simplified if we consider translationally invariant systems, which play a fundamental role in statistical physics. In this case the system is fully described by the elementary cell, and we can state the following.\vs

\noindent {\em A translationally invariant  system obeys conditions (2a-b) if and only if they are satisfied for the elementary cell}.\vs

The {\em only if} implication follows naturally. For the {\em if} implication consider two adjacent elementary cells, $\mathcal C_1$ and $\mathcal C_2$ [Fig.~\ref{fig:lattice}]. Condition {\em (2a)} is  automatically satisfied. To see that {\em (2b)} is also satisfied let $a\in\mathcal C_1$ and $b\in \mathcal C_2$, and consider two paths $p_1$, $p_2$ where $p_i(a\rightarrow   b)$ goes from $a$ to $b$ passing through an intermediate site $c_i\in \mathcal C_1 \cap \mathcal C_2$. Then
\begin{align}\nonumber
	\pi(p_1(a\rightarrow   b))%=&\,\pi(a\rightarrow   c_1)\pi(c_1\rightarrow   b)\\
	\nonumber
	=&\, \pi(a\rightarrow   c_1)[\pi(c_1\rightarrow   c_2)]^2\pi(c_1\rightarrow   b)\\
\nonumber
	=&\, \pi(p_2(a\rightarrow   b)).
\end{align}
The proof is concluded by realizing that this argument trivially extends to other elementary cells.\hfill$\Box$

Remarkably, a generalized form of Toulouse criterion for testing conditions {\em(2a-b)} on the elementary cell can be derived.\vs

\noindent{\em If a model satisfies
\begin{equation}\label{ToulouseVec.app}
	\pi(\ell)=+\openone.
\end{equation}
for every loop $\ell$ in the elementary cell and all coupling vectors have the smallest component along the same axis, then it is an INES model}.\vs

If conditions {\em 2a-b)} are satisfied then the model is an INES model. For translationally invariant systems, it suffices that the conditions hold within the elementary cell. Hence, it only remains to prove that Eq.~(6) is necessary and sufficient for the elementary cell to obey {\em 2a-b)}.

Any two paths $p_1(a\rightarrow   b)$, $p_2(a\rightarrow   b)$ can be merged into a closed loop $\ell=p_1(a\rightarrow   b)\cup p_2(a\rightarrow   b)$, hence {\em 2b)} implies $\pi(\ell)=[\pi(a\rightarrow   b)]^2=+1$. On the other hand, if any closed loop yields $\pi(\ell)=+1$, then any two paths $p_1(a\rightarrow   b)$, $p_2(a\rightarrow   b)$ must have $\pi(p_1)=\pi(p_2)$.\hfill$\Box$

\section{Numerical evidence of the sufficient conditions (Conjectures 1 and 2)}
In order to test Conjectures 1 \& 2 we have generated several Hamiltonians without geometric frustration, and checked that they satisfy the equality $f=E$ in Eq.(3) in the main text for all pairs of spins. Instances of the prototype model have been generated, where the gauge was fixed for the unitary sector of the gauge group, since this symmetry is trivially proven analytically. In order to check Conjecture 2, a random gauge transformation consisting of pure parity transformations was applied. In total 278400 geometric frustration-free Hamiltonians have been checked by exact diagonalization. In all those without global ground-state degeneracy, no instance was found which failed to satisfy $f=E$ for all pairs of spins.

In the following we provide more details of our procedure. We have evaluated 278400 spin-1/2 models on random graphs with $n=4,6$ and $8$ spins (sites). All the numerical tests have been done using the program MATHEMATICA 6. We have generated random models of the $XYZ$ type (general anisotropic Heisenberg), $XXZ$ and $XXX$ (Heisenberg) satisfying conditions {\it 2a-b)}, both with homogeneous and inhomogeneous couplings.
The models have been generated according to the following procedure. First, a connectivity graph $(\mathcal V,\mathcal E)$ with $|\mathcal V|=n$ vertices connected by edges $\mathcal E$ is generated. For each edge $(ij)\in\mathcal E$, an interaction of the types $XYZ$, $XXZ$ or $XXX$ was generated. Depending on whether an homogeneous or inhomogeneous model was being generated, the interactions were obtained independently, or the same interaction was generated for all edges respectively. Finally, the identity or a parity inversion was applied to each spin, randomly with probability $1/2$.
%This the gauge was Þxed for the unitary sector of the gauge groupof 0.1.
At each value of the edge probability $q$, and for each type of model, the
method exhaustively evaluates all possible connected simple graphs with $n$ vertices.
Specifically, given a simple connected graph $(\mathcal V,\mathcal E)$, the resulting Hamiltonians are, for the homogeneous case
\begin{equation}
	H_{\textrm{hom}}=-\sum_{\nu\in\{x,y,z\}}J^\nu\sum_{(ij)\in \mathcal E}(-)^{x_i\oplus x_j}S_i^\nu S_j^\nu,\\
\end{equation}
and for the inhomogeneous case,
\begin{equation}
	H_{\textrm{inhom}}=-\sum_{\nu\in\{x,y,z\}}\sum_{(ij)\in \mathcal E}(-)^{x_i\oplus x_j}J_{ij}^\nu S_i^\nu S_j^\nu,\\
\end{equation}
where the $x_i$ are equiprobable binary random variables and each interaction was obtained by generating a ferromagnetic coupling vector $\vec J_{ij}$ with $J_{ij}^x,J_{ij}^y$ and $J_{ij}^z$, each component according to a flat p.d.f. (probability distribution function) [$dp(J)=dJ$] between $0$ and $1$ for the general $XYZ$ case. Ensuring condition {\it 2a)} was achieved by placing the smallest component on the $z$ axis, and leaving the other two components in their original order. Notice that the choice of the $z$ axis for the smallest component is arbitrary and the two other choices are equivalent up to a global $SU(2)$ transformation. For the $XXZ$ model the $J^x$ and $J^y$ components were symmetrized by replacing the obtained values by their averages. Finally, the $XXX$ interactions were obtained by simply generating random numbers between $0$ and $1$.

The connectivity graphs were obtained as follows. For $n=4$ and $n=6$, we tested all connected simple graphs with $n$ vertices. For $n=8$, and due to the very large number of inequivalent connected simple graphs (11117), we iterated the Erdos-R\'enyi model~\cite{erdosrenyi} to generate simple connected random graphs.
In order to test all different kinds of connectivity (ranging from systems with few interactions such as the one-dimensional spin chain up to the fully connected model), the edge probability $q$ was run from $0.1$ to $1$ with increments of $0.1$ steps. At each value of $q$, and  for each type of model ($XYZ$, $XXZ$ or $XXX$), $2000$ instances were generated, totaling at $18000$ models for each type of interaction.

In Tables~\ref{table1}, \ref{table2} and \ref{table3} we present our results, showing the total number of models generated for each type of interaction considered. The number of {\em accepted} Hamiltonians (those for which the condition $f=E$ is satisfied for all bonds $(ij)\in\mathcal E$), {\em rejected} Hamiltonians (those violating $f=E$ for some bond) and {\em degenerate} Hamiltonians (those with globally degenerate ground state) are shown for each interaction class. As shown by the tables, no Hamiltonians have been {\em Rejected}, meaning that no instance has been found violating conjectures 1 \& 2. This finding lends strong support to the general validity of the conjectures.

\begin{table}[b]
\begin{center}
\begin{tabular}{ccccc}
$n$&~~{\#} graphs~~& Accepted&Rejected&Degenerate\\
\hline
\hline
4&6&6000&0&0\\
\hline
6&112&22400&0&0\\
\hline
8&11117&18000&0&0\\
\hline
\\
\hline\hline
4&6&6000&0&0\\
\hline
6&112&22400&0&0\\
\hline
8&11117&18000&0&0\\
\hline
\end{tabular}
\end{center}
\caption{\label{table1} Results for $XYZ$ models with homogeneous (above) and inhomogeneous (below) interactions.}
\begin{center}
\begin{tabular}{ccccc}
$n$&~~{\#} graphs~~& Accepted&Rejected&Degenerate\\
\hline
\hline
4&6&6000&0&0\\
\hline
6&112&22400&0&0\\
\hline
8&11117&18000&0&0\\
\hline
\\
\hline\hline
4&6&6000&0&0\\
\hline
6&112&22400&0&0\\
\hline
8&11117&18000&0&0\\
\hline
\end{tabular}
\end{center}
\caption{\label{table2} Results for $XXZ$ models with homogeneous (above) and inhomogeneous (below) interactions.}
\begin{center}
\begin{tabular}{ccccc}
$n$&~~{\#} graphs~~& Accepted&Rejected&Degenerate\\
\hline
\hline
4&6&2247&0&3753\\
\hline
6&112&7072&0&15328\\
\hline
8&11117&4902&0&13098\\
\hline
\\
\hline\hline
4&6&2262&0&3738\\
\hline
6&112&7088&0&15312\\
\hline
8&11117&4908&0&13092\\
\hline
\end{tabular}
\end{center}
\caption{\label{table3} Results for $XXX$ models with homogeneous (above) and inhomogeneous (below) interactions.}
\end{table}%


\begin{thebibliography}{99}

\bibitem{Toulouse1977} G. Toulouse, Commun. Phys. {\bf 2}, 115 (1977).

\bibitem{Villain1977} J. Villain, J. Phys. C {\bf 10}, 1717 (1977); S. Kirkpatrick, Phys. Rev. B {\bf 16}, 4630 (1977).

\bibitem{Binder1986} K. Binder and A. P. Young, Rev. Mod. Phys. {\bf 58}, 801 (1986); H. T. Diep (ed.), {\it Frustrated spin systems}, World Scientific, Singapore (2005).

\bibitem{Wolf2003} M. M. Wolf, F. Verstraete, and J. I. Cirac, Int. J. Quant. Inf. {\bf 1}, 465 (2003).

\bibitem{Nielsen2004} C. M. Dawson and M. A. Nielsen, Phys. Rev. A {\bf 69}, 052316 (2004).

\bibitem{Giampaolo2010} S. M. Giampaolo, G. Adesso, and F. Illuminati, Phys. Rev. Lett. {\bf 104}, 207202 (2010).

\bibitem{Eisert2010} N. de Beaudrap, M. Ohliger, T. J. Osborne, and J. Eisert, Phys. Rev. Lett. {\bf 105}, 060504 (2010).

\bibitem{Acin2007} A. Ferraro, A. Garc\'{i}a-Saez, and A. Acin, Phys. Rev. A {\bf 76}, 052321 (2007).

%%\bibitem{Eisert2} N. de Beaudrap, T. J. Osborne, J. Eisert, {\it New J. Phys.} {\bf 12} 095007 (2010).
%%\bibitem{Melko} R. G. Melko, A. Paramekanti, A. A. Burkov, A. Vishwanath, D. N. Sheng, L. Balents, {\it Phys. Rev. Lett.} {\bf 95}, 127207 (2005).

\bibitem{Facchi2010} P. Facchi, G. Florio, U. Marzolino, G. Parisi, and S. Pascazio, New J. Phys. {\bf 12}, 025015 (2010).

\bibitem{Lacorre} P. Lacorre, J. Phys. C {\bf 20}, L775 (1987); A. Sen De, U. Sen, J. Dziarmaga, A. Sanpera, and M. Lewenstein, Phys. Rev. Lett. {\bf 101}, 187202 (2008).

%%\bibitem{McKenzie} W. Zheng {\it et al}, {\it Phys. Rev. B} {\bf 71}, 134422 (2005).

\bibitem{Balents} L. Balents, Nature Phys. {\bf 464}, 199 (2010), and references therein.

\bibitem{Ramirez} A. P. Ramirez, Annu. Rev. Mater. Sci. {\bf 24}, 453 (1994).

\bibitem{Sengstock} J. Struck, C. \"Olschl\"ager, R. Le Targat, P. Soltan-Panahi, A. Eckardt,
M. Lewenstein, P. Windpassinger, and K. Sengstock, Science {\bf 333}, 996 (2011).

\bibitem{Greiner} J. Simon, W. S. Bakr, R. Ma, M. E. Tai, P. M. Preiss, and M. Greiner,
Nature {\bf 472}, 307 (2011).

\bibitem{Monroe} K. Kim, M.-S. Chang, S. Korenblit, R. Islam, E. E. Edwards, J. K. Freericks, G.-D. Lin,
L.-M. Duan, and C. Monroe, Nature {\bf 465}, 590 (2010).

\bibitem{Plenio} A. Bermudez, J. Almeida, F. Schmidt-Kaler, A. Retzker, and M. B. Plenio,
arXiv:1108.1024 (2011).

\bibitem{Verstraete} V. Murg, F. Verstraete, and J. I. Cirac, Phys. Rev. B {\bf 79}, 195119 (2009).

\bibitem{Turgut} K. Uyan\ifmmode \imath \else \i \fi{}k and S. Turgut, Phys. Rev. A {\bf 81}, 032306, (2010).

\bibitem{Dellanno} M. Blasone, F. Dell'Anno, S. De Siena, and F. Illuminati, Phys. Rev. A {\bf 77}, 062304 (2008).

%%\bibitem{generalmon} J. S. Kim, A. Dan, B. C. Sanders, {\it Phys. Rev. A} {\bf 79}, 012329 (2009).

\bibitem{CKW} V. Coffman, J. Kundu, and W. K. Wootters, Phys. Rev. A {\bf 61}, 052306 (2000); T. J. Osborne and F. Verstraete, Phys. Rev. Lett. {\bf 96}, 220503 (2006).

\bibitem{Partialtranspose} A. Peres, Phys. Rev. Lett. {\bf 77}, 1413 (1996); M. Horodecki, P. Horodecki, and R. Horodecki, Phys. Lett. A {\bf 223} 1, (1996);  R. Horodecki, P. Horodecki, M. Horodecki, and K. Horodecki, Rev. Mod. Phys. {\bf 81}, 865 (2009).

\bibitem{MERA} G. Vidal, Phys. Rev. Lett. {\bf 99}, 220405 (2007).

\bibitem{Pyrochlore} A. Koga and N. Kawakami, Phys. Rev. B {\bf 63}, 144432 (2001).

\bibitem{Vidal2010} G. Evenbly and G. Vidal, Phys. Rev. Lett. {\bf 104}, 187203 (2010).

\bibitem{Wolf2008} M. M. Wolf, Phys. Rev. Lett. {\bf 100}, 070502 (2008); J. Eisert, M. Cramer, and M. B. Plenio, Rev. Mod. Phys. {\bf 82}, 277 (2010).

\bibitem{Monras2011} A. Monras, G. Adesso, S. M. Giampaolo, G. Gualdi, G. B. Davies, and F. Illuminati, Phys. Rev. A {\bf 84}, 012301 (2011).

%\bibitem{Review} R. Horodecki, P. Horodecki, M. Horodecki, and K. Horodecki, Rev. Mod. Phys. {\bf 81}, 865 (2009).

%%\bibitem{HierarchyCollective} F. Dell'Anno, S. M. Giampaolo, S. De Siena, and F. Illuminati, %%arXiv:1103.yyyy (2011).

%%\bibitem{LocalUnitaries} A. Monras, G. Adesso, S. M. Giampaolo, G. Gualdi, and F. Illuminati, %%arXiv:1103.zzzz (2011).

%%\bibitem{Bhatia} R. Bhatia, {\it Matrix analysis}, Graduate Texts in Mathematics, vol. {\bf 169}, Springer, %%New York, 1996.

%%\bibitem{Markham} E. Fu and T. L. Markham, {\it Linear Algebra Appl.} {\bf 179}, 7 (1993).

\end{thebibliography}

\begin{thebibliography}{99}

\bibitem{Bhatia} R. Bhatia, {\it Matrix analysis}, Graduate Texts in Mathematics, vol. {\bf 169}, Springer, New York, 1996.

\bibitem{Markham} E. Fu and T. L. Markham, {\it Linear Algebra Appl.} {\bf 179}, 7 (1993).

\bibitem{erdosrenyi} E.~N.~Gilbert, Ann. Math. Stat. {\bf 30}, 1141 (1959). P. Erdos and A. R\'enyi, Publ. Math. (Debrecen) {\bf 6}, 290 (1959). P. Erdos and A. R\'enyi, Publ. Math. Inst. Hung. Acad. Sci. {\bf 5}, 17 (1960).
\end{thebibliography}
\end{document}